\documentclass[useAMS,usenatbib]{mn2e}
\usepackage{epsfig}
\usepackage[usenames,dvips]{color}
\usepackage{amsmath}

\newcommand{\Msolar}{\mbox{\,$\rm M_{\odot}$}} % solar mass
\def\simge{\mathrel{\raise1.16pt\hbox{$>$}\kern-7.0pt \lower3.06pt\hbox{{$\scriptstyle\sim$}}}} % approx ge
\def\simle{\mathrel{\raise1.16pt\hbox{$<$}\kern-7.0pt \lower3.06pt\hbox{{$\scriptstyle \sim$}}}} % approx le

\begin{document}

\title[NSC-FRB system]{What binary systems are the most likely sources for periodically repeating FRBs?}
\author[]{Xianfei
Zhang$^{1}$\thanks{E-mail: zxf@bnu.edu.cn} and He Gao$^{1}$\thanks{E-mail: gaohe@bnu.edu.cn}
\\ $^1$ Department of Astronomy, Beijing Normal University, Beijing 100875, China}

%\linespread{2.0} \begin{document}
\date{Accepted . Received ; in original form }

\pagerange{\pageref{firstpage}--\pageref{lastpage}} \pubyear{2011}

\maketitle

\label{firstpage}

\begin{abstract} The newly discovered 16.35 days period for repeating FRB 180916.J0158+65 provides an essential clue for understanding the sources and emission mechanism of repeating FRBs. Many models propose that the periodically repeating FRBs might be related to binary star systems that contain at least one neutron star (NSC-FRB system). It has been suggested that the NS ``combed" by the strong wind from a companion star might provide a solution. Following the ``Binary Comb" model, we use the population synthesis method to study in detail the properties of the companion stars and the nature of NSC-FRB systems. Our main findings are:  1) the companion star is most likely to be a B-type star; 2) the period of 16 days of FRB 180916 happens to fall in the most probable period range, which may explain why FRB 180916 was the first detected periodically repeating FRB, and we expect to observe more periodically repeating FRBs with periods around 10-30 days; 3) the birth rate for NSC-FRB system is large enough to fulfill the event rate requirement set by the observation of FRB 180916, which supports the proposal that NSC-FRB can provide one significant channel for producing periodically repeating FRBs.
\end{abstract}


\section{Introduction}
Fast radio bursts (FRBs) have been extensively explored since their discovery more than 10 years ago \citep{Lorimer2007,Thornton2013}, but the nature of the FRBs' sources remains a mystery \cite[][for reviews]{Petroff2019,CordesandChatterjee2019}. In the literature, FRBs are usually classified into two categories: non-repeating FRBs which are proposed to be related to cataclysmic events, such as binary neutron star mergers \citep{Totani2013,Wang2016,Yamasaki2017,DokuchaevandEroshenko2017,wang20}, binary white-dwarf mergers \citep{Kashiyama2013}, mergers of neutron star and black hole \citep{zhang19,dai19} or mergers of charged black holes \citep{Zhang2016b,Liu2016}, collapses of supramassive rotating neutron stars \citep{FalckeandRezzolla2013,Zhang2014,Ravi2014,PunslyandBini2016} and so on; or repeating FRBs which might be related to some repeatable astrophysical phenomenons, such as magnetar flares \citep{Kulkarni2014,Lyubarsky2014}, collisions and interactions between neutron stars and small objects \citep{GengandHuang2015,Dai2016,MottezandZarka2014,Smallwood2019}, quark novae \citep{Shand2016}, giant pulses of pulsars \citep{Connor2016,CordesandWasserman2016}, cosmic combs \citep{Zhang2017,Zhang2018} and so on. See \cite{platts18} for a review on the available theoretical models.

Most recently, Canadian Hydrogen Intensity Mapping Experiment (CHIME) telescope reports that their first discovered repeating FRB 180916.J0158+65 (hereafter FRB 180916) exhibits an activity period of 16.35 days with 4.0-days active time window \citep{CHIME20}. This may bring clues for understanding the source and emission mechanism of repeating FRBs. Up to now, some models have been proposed to interpret the periodic behavior of FRB 180916, and most of them introduce the orbital periods of a neutron star (NS) and its companion star to explain the periodicity \footnote{\cite{yang20} proposed that the periodicity of FRB 180916 comes from orbital-induced, spin precession of the FRB emitter (most likely a NS), where the companion is likely a compact star and the binary period ranges from several hundreds to thousands of seconds.}. For instance, \cite{ioka20} proposed that FRB 180916 is produced by a highly magnetized pulsar with a lifetime smaller than $10^4$ yr, whose magnetic field is ``combed" by the strong wind from a companion star. And they considered two situations that the companion star could be either a massive star or a millisecond pulsar. \cite{lyutikov20} also considered to use NS and massive star binary to interpret the periodicity of FRB 180916, but they proposed that the observed periodicity is due to the orbital phase-dependent modulation of the
absorption conditions in the massive star's wind, instead of intrinsic to the FRBs' source. \cite{dai20} propose that periodically repeating FRB might  be produced from an old-aged pulsar traveling through the asteroid belts of its companion star, so that the periodically repeating FRBs would provide a unique probe of extragalactic asteroid belts. \cite{gu20} suggested that FRB 180916 is produced by a NS-white dwarf (WD) binary system when the WD fills its Roche lobe at the pericenter, so that the mass transfer from the WD to the NS resulting in multiple radio bursts. In this case, the observed period of FRB is identical to the orbital period of the binary.

Under the ``Binary Comb" model framework proposed by \cite{ioka20}, this \emph{Letter} tries to investigate the following questions by population synthesis method: when a binary star system evolves into the ``FRB-productive" stage, namely at least one of the stars has evolved into a NS with lifetime being less than $10^4$ yr and the companion's wind density is not opaque to induced Compton or Raman scatterings for repeating FRB emission (henceforth, we call such a system as NSC-FRB system), 1) what are the properties of its companion star, e.g., surface temperature, luminosity and masses; 2) what is the distribution of the orbital period and eccentricity, is the 16 day period unique across the distribution; 3) what is the birth rate for NSC-FRB systems, is it large enough to fulfill the event rate requirement set by the observation of FRB 180916 ?

%%========================================================
\section{The models}
\label{s_models}

For the purpose of this work, we first simulate $10^7$ systems including single and binary stars, and evolve them starting from zero-age main-sequence stage within an evolution time of $14\,\rm{Gyr}$. Here we calculate the evolution models of binary stellar stars, including individual stellar evolution tracks and initial conditions, with the \texttt{BSE} code \citep{Hurley00,Hurley02}. Some of the important initial conditions and physical processes are described in the following. These basic parameters for binary star evolution chosen in this work have been used in several previous studies, e.g., for neutron star formation \citep{demink2015,Zapartas2017,wang20}.\\

\noindent (1) We adopt the initial mass function (IMF) of \citet{Kroupa2001} in the range $0.1-100\,\Msolar$.
The initial mass of primary stars (the massive star in system), $M_1$, are generated by the formula of
\begin{equation}\label{eq:IMF}
 \frac{\mathrm{d}N}{\mathrm{d}M_1} \propto M_{1}^{\alpha} ,
\end{equation}
where $\alpha = -2.3$.

\noindent (2) We assume a mass-ratio distribution of
\begin{equation}
\frac{\mathrm{d}N}{\mathrm{d}q} \propto q^{\kappa}.
\label{eq:iqf}
\end{equation}
We adopt $\kappa = 0$ which consistent with \citet{Sana2012} and \citet{Kiminki2012}.

\noindent (3) We assume a distribution of \textbf{orbital periods} by the formula of
\begin{equation}
\frac{\mathrm{d}N}{\mathrm{d}\log_{10}P} \propto \left( \log_{10} P \right)^{\pi}.
\end{equation}
We adopt $\pi = 0$ for $M_1 \le 15\Msolar$ \citep{Kobulnicky2014,Moe2015} and $\pi = -0.55$ for $M_1 > 15\Msolar$ \citep{Sana2012}.

\noindent (4) We assume the distribution of initial eccentricities of binaries follows
\begin{equation}
f_e (e) \propto  e^\eta.
\end{equation}
We adopt $\eta = 0.42$ as found by \citet{Sana2012}.

\noindent (5) The observations indicate that the binary fraction may depend on
the binary parameters \citep{Kouwenhoven2009,Sana2012}. Hence, we assume a binary fraction by the formula of \citet{vanhaaften2013}
\begin{equation}
f_b=0.5+0.25\,\rm log_{10}(M_1).
\label{eq:fb}
\end{equation}

\noindent (6) The NS have kick velocities due to binary evolution \citep{Iben96} and
explosion of supernovae \citep{Lai95,Lai2001,Nordhaus2012}.
We use the formula of \citet{Hansen97}, a Maxwellian distribution, and  Monte Carlo
procedure to generate the individual kick velocities for the neutron stars:
\begin{equation}
\frac{{\rm d}N}{N{\rm d}v}=\left(\frac{2}{\pi}\right)^{1/2}\left(\frac{v^{2}}{\sigma^{3}}\right)e^{-v^{2}/2\sigma^{2}},
\end{equation}
where $v$ is the kick velocity and $\sigma$ is its dispersion, ${\rm d}N/N$ is the normalized number
in a kick velocity bin ${\rm d}v$, and $\sigma=190$ km s$^{-1}$ based on analysis of pulsar samples \citep{Hansen97}. The kick velocities will affect the
eccentricity of systems during the evolution.

\noindent (7) Considering that the spiral host galaxy of FRB 180916 is similar to our Milky Way in mass and metallicity \citep{CHIME20},
here we adopt a metallicity of Z=0.02 in our calculation.

\noindent (8) Once the Roche lobe mass transfer between the NS and its companion star happens, the mass transfer rate is relatively large so that the optical depth for most cases would become too large to form an FRB within the ``Binary Comb" model. Therefore, here we only consider the binary systems without Roche lobe mass transfer.

\begin{figure*}
\includegraphics{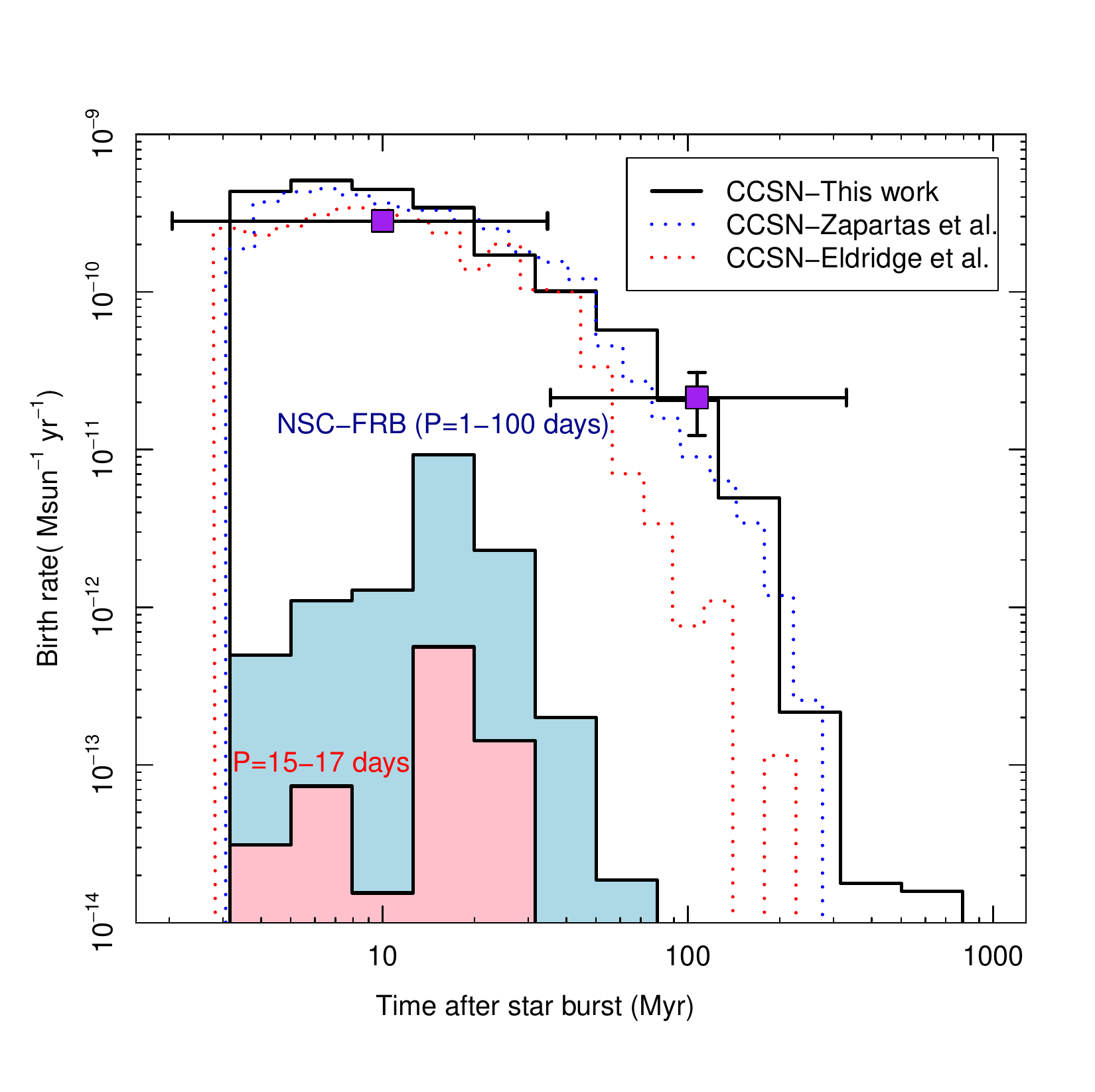}
\caption{The birth rate of CCSN and NSC-FRB system. The solid lines are birth rate of CCSN, P=1-100 days NSC-FRB systems and
P=15-17 days NSC-FRB systems in our results, respectively. The dotted lines are theoretic birth rate of CCSN from \citet{Zapartas2017} and \citet{Eldridge2019}, respectively.
The pink squares shows the observation value for CCSN from Magellanic Cloud supernova remnants\citep{Maoz2010}. Note that, the results of \citet{Maoz2010} cannot easily distinguish between CCSN and possible type Ia supernovae. For the left pink square, it is CCSN occur at such low delay-times, but the right pink square may include both types of supernovae. }
\label{rate}
\end{figure*}

The outcomes of stellar evolution are following the same definition in \citet{Hurley00,Hurley02}, e.g., NS stars are formed from massive star ($M \ge 7 \Msolar$) core-collapse supernovae (CCSN). Of the $10^7$ systems, about $55\%$ initial zero-age systems are binaries which determined by Eq.~\ref{eq:fb}.
By considering the formation channel of NS from core-collapse supernovae, we compare our results to other theoretical works and the observations. As shown in Fig. \ref{rate}, the birth rate of NS at different delay-time (time after starburst) from our work are full consistent with previous works and also match the observation values.

Furthermore, we screen out those systems that can generate FRBs according to the ``Binary Comb model". Here we set two criteria to become a NSC-FRB system: \\
(1) a relative young NS ($t_{\rm life}\le 10^4 \rm yr$), which have enough energy and enough luminosity for pertaining a funnel in the
wind from the companion; \\
(2) a suitable optical depth ($\tau_{\rm C}<10$, \citet{Lyubarsky2008}) to the induced Compton scattering. Here we check the criterion at the maximum binary separation distance $r=a(1+e)$, where $a$ and $e$ are the semi-major axis and the eccentricity of the binary, respectively. We estimate $\tau_{\rm C}$ by (the same as Eq.4 of \cite{ioka20})
\begin{eqnarray}
\tau_{C} &\sim& \frac{3\sigma_T}{32\pi^2}
  \frac{n_w(r) L c \Delta t}{r^2 m_e \nu^3}
  \sim 30\, {\dot M}_{-9} V_{3.3}^{-1} r_{13}^{-4} (L\Delta t)_{38} \nu_9^{-3},
\label{eq:tao}
\end{eqnarray}
where $\sigma_{T}$ is Thomson cross section, $L \Delta t = 10^{38}\,{\rm erg}\,(L\Delta t)_{38}$ is adopted as the FRB isotropic luminosity times duration and $\nu=1$\,GHz\,$\nu_9$ is adopted as the fiducial frequency. Similar to \cite{ioka20}, here we estimate the wind density $n_w(r)$ around the NS as $n_w(r) \sim n_w(0) [a/(a+r)]^{2}$, where
\begin{eqnarray}
n_w(0)\sim  \frac{\dot{M}}{4 \pi a^2m_pV}.
\label{eq:nw0}
\end{eqnarray}
$V$ is the wind velocity, and $m_p$ is the proton mass. Here we adopt $V\sim2\times10^3 \rm~km~s^{-1}$. It is worth noticing that this is a typical estimation for massive main-sequence stars but not for giant stars. The wind velocity of a giant star could be 1-2 order lower than a massive main-sequence star. Considering that there are very few NS+giant star systems existing in our final sample, here we simply apply this value to all systems, which should not affect the final results. The mass-loss rate ${\dot M}=10^{-9} M_{\odot}\,{\rm yr}^{-1} {\dot M}_{-9}$ is due to stellar winds, we selected parameters similar to the MESA isochrones and stellar tracks (MIST) project for low mass and massive stars \citep{Dotter2016,Choi2016}. In MIST, the mass loss rate of stars are treated via different formula, i.e.,
Reimers stellar winds \citep{Reimers75,McDonald2015} for low mass stars ($M<10\Msolar$) and the formulas of \citet{vink00,vink01} for massive stars. On the other hand, when the companion star is another NS, the wind density around the FRB source is \citep{ioka20}
\begin{eqnarray}
  n_{w}(0) &\sim& \frac{L_w}{4\pi a^2 m_e c^2 V \Gamma (1+\sigma)}
\end{eqnarray}
where $L_{w}$ is the pulsar wind luminosity, $\sigma$ is the ratio of Poynting flux to particle energy flux, and $\Gamma=[1-(V/c)^2]^{-1/2}$ is the Lorentz factor of the wind.

%------------------------------------------------------------------------------

\section{Results}
\label{s_results}

\begin{figure*}
\includegraphics{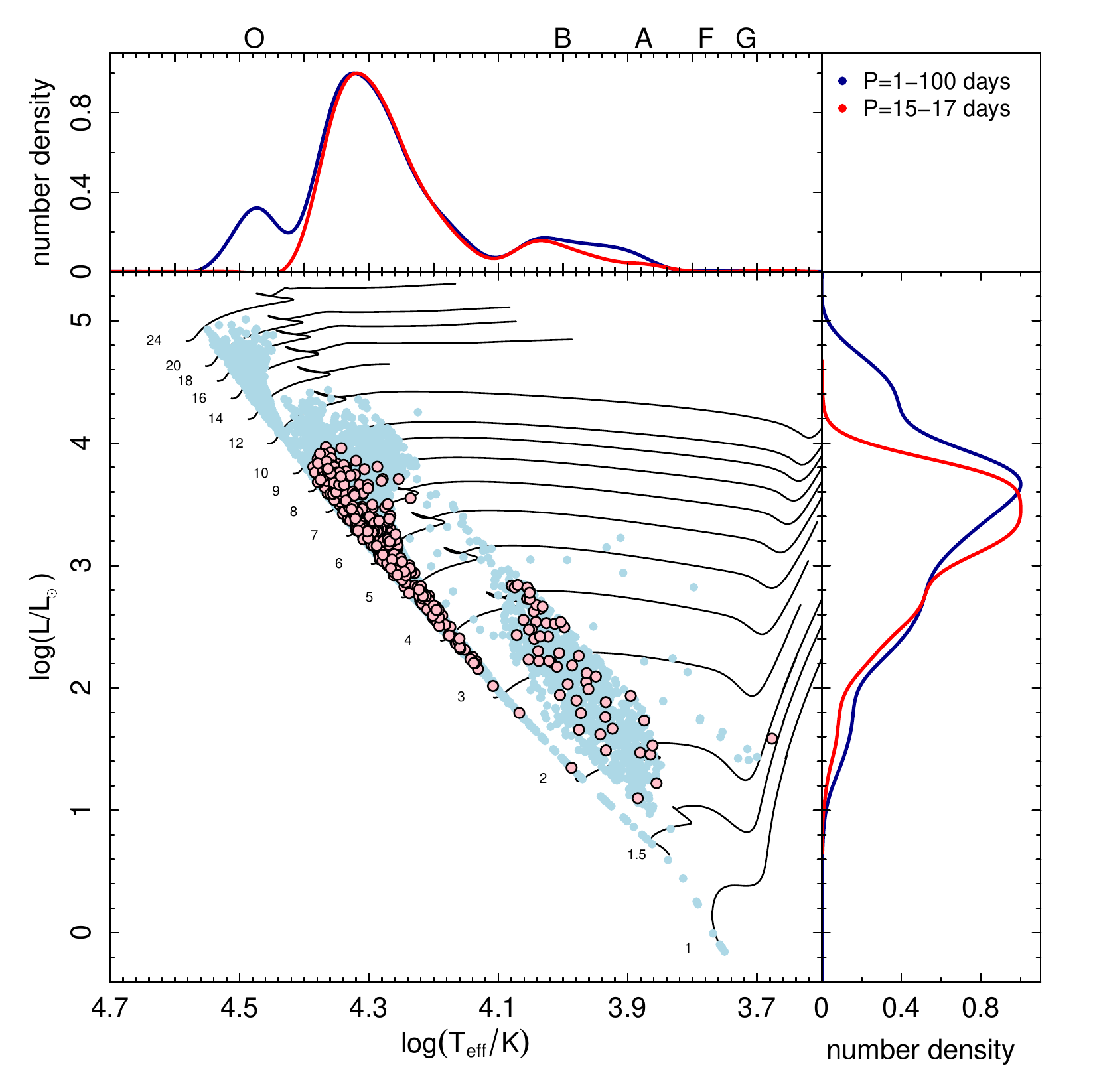}
\caption{The HR diagram of the MS companion stars.The red dots and blue dots indicate the P=15-17 days NSC-FRB systems and
the P=1-100 days NSC-FRB systems,respectively. In the top-left and lower-right panels, curves with different colours represent the normalized number density distributions of the NSC-FRB systems in $T_{\rm eff}$ and $\log L$ respectively. }
\label{hr}
\end{figure*}

\begin{figure*}
\includegraphics{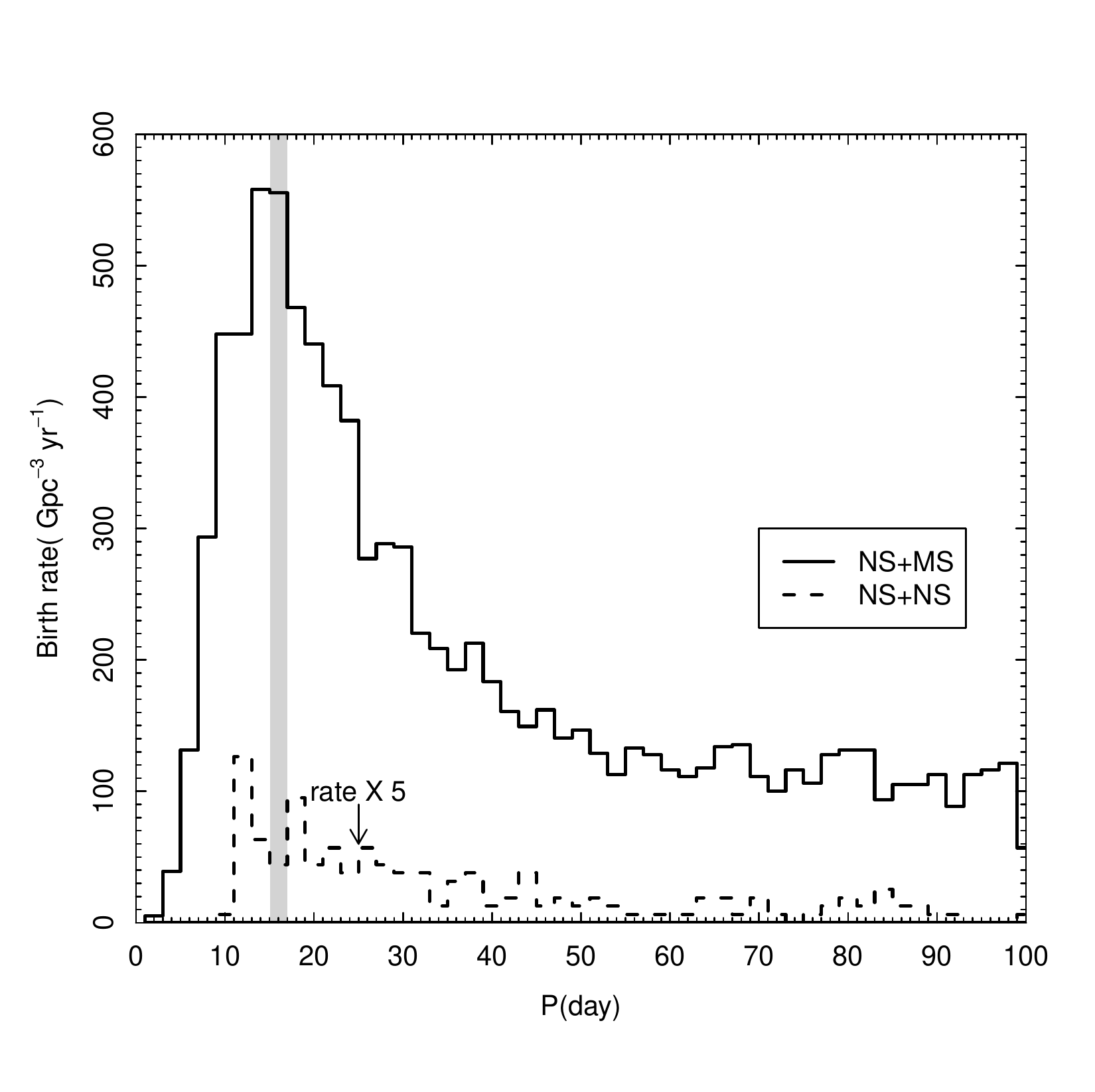}
\caption{Distribution of volumetric birth rate for NS+MS and NS+NS FRB systems. The birth rate of DNS has increased to 5 times for a clear view.
The grey zone indicate the approximate period range of 15-17 days for FRB 180916.}
\label{events}
\end{figure*}

Eventually we obtain $7808$ NSC-FRB systems in the periods of 1-100 days\footnote{As shown in Fig. \ref{events}, the birth rate of NSC-FRB systems with periods outside this interval significantly decays.}. The properties for the system and for the companion stars are summarized as follows:
\begin{itemize}
\item $96.7\%$ ($7554/7808$) of the companion stars are core hydrogen-burning main-sequence stars, $0.3\%$ ($21/7808$) are in shell hydrogen-burning stage, $0.7\%$ ($60/7808$) are helium stars and $2.3\%$ ($173/7808$) are neutron stars. Since the NSC-FRB system requires a newly born NS with lifetime much shorter than the time to produce white dwarfs, there is none NS+WD type system in our sample. In short, the dominant type of NSC-FRB system is NS + MS binary, which are much more than other types of binaries.
\item Fig.~\ref{hr} shows all the MS companion stars (the main population of NSC-FRB systems in our sample) in the Hertzsprung-Russell diagram. We find that most of the stars are on F to O type, and few ($<0.2\%$) are G type. The dominant type of stars is B-type ($>85\%$). The mass of MS companion stars are in range of 1-24 \Msolar, with a peak between 4 and 12 \Msolar.
\item Considering that the period of FRB 180916 is around 16 days, NSC-FRB systems with a period of 15 to 17 days was analyzed separately. There are $446$ NSC-FRB systems with the period in a range of 15-17 days. For these systems, the companion stars are $97.7\%$ ($436/446$) core-burning main-sequence stars, $1/446$ ($0.2\%$) in shell hydrogen-burning stage, $0.4\%$ ($2/446$) helium stars and $1.7\%$ ($7/446$) neutron stars. Among the MS stars, the dominant type is B-type ($>85\%$), and the star masses are in range of 1.5-10 \Msolar, with a peak between 4 and 10 \Msolar.
\item The ages of most NSC-FRB systems are younger than 50 Myr. There is a relatively wide eccentricity distribution from 0.2 to 0.9, with a flat peak between 0.5-0.9 ($\sim 80\%$ of systems), where the upper boundary mainly comes from the limitation of no Roche lobe overflow. Most interestingly, the orbital period distribution has a relatively sharp peak just around  16 days, and $43.5\%$ systems have period within 10-30 days.
\end{itemize}
Based on our results, we can judge from the probability that if the periodically repeating FRBs are indeed produced from NSC-FRB systems, the most likely companion of the neutron star is B-type main-sequence star. For a specific sample with FRB 180916 like period ($\sim 16$ days), the conclusion remains unchanged. NSC-FRB system is likely to be close to the star formation regions, which is consistent with the observation of FRB 180916 \citep{CHIME19,CHIME20}. The period of 16 days happens to fall in the most probable period range, which may explain why FRB 180916 was the first periodically repeating FRB to be detected. If our interpretation is correct, we would expect to observe more periodically repeating FRBs with periods similar to FRB 180916, or possible widely in 10-30 days.

Finally, we give a ballpark estimation for the absolute birth rate of NSC-FRB systems with different orbital period. Considering that FRB 180916 located in a  Milky Way-like spiral galaxy, in our calculation we assume a constant star formation rate of $3.5\,\Msolar\,\rm{yr}^{-1}$ according to the recent Milky Way value; \citep{Diehl2006,Dominik2012,Fantin2019}, and assume a constant galaxy number density of $\sim 10^6 \rm Gpc^{-3}$, which is the local cumulative galaxy number density corresponding to the Milky Way mass objects \citep{torrey15}. Hence, we can convert the birth rates for NSC-FRB systems in different delay-time (Fig.~\ref{rate}) into an approximate volumetric birth rate, as used by the LIGO/Virgo collaboration, with the formula of (see also in \citet{demink2015})
\begin{equation}
{\cal R}_{\rm vol} =  10 \ {\rm yr}^{-1} \ {\rm Gpc}^{-3}\
\left[\frac{\rho}{0.01 \rm Mpc^{-3}}\right]\left[\frac{ {\cal R}_{\rm gal}}{\rm Myr^{-1}}\right]	
\end{equation}
where $\rho$ is the local density of Milky Way-like galaxies and $R_{\rm gal}$ is the
birth rates for NSC-FRB systems in Milky Way-like galaxy. Fig.~\ref{events} shows the distribution of volumetric birth rate along with different periods. Here we plot the birth rate of NS+MS and NS+NS systems separately. The birth rate of NS+MS systems is more than 10 times of NS+NS systems, which proves again that if periodically repeating FRBs are indeed produced by NSC-FRB systems, then most of them comes from NS+MS (mostly likely NS-B-type star) systems. For NS+MS systems in 15-17 days, the birth rate would be $5.5 \times 10^{2}\,\rm Gpc^{-3} yr^{-1}$, which is about ten times of the source birth rate estimated by \cite{ioka20} ($\sim 30\,\rm Gpc^{-3} yr^{-1}$) based on the FRB observations (especially the properties of FRB 180916). The difference is understandable, since in our calculation, we did not consider the restrictions for the magnetic field of the newly born NS ($B_{\rm p,NS}$). According to the ``Binary Comb model", for MS companion cases, $B_{\rm p,NS}$ should be larger than $10^{13}$ G \citep{ioka20}. The magnetic field distribution of newly born NSs is still very uncertain. Based on the Milky Way observations, \cite{Beniamini19} argues that a fraction (in order of $0.1$) of NSs being born as highly magnetized, and their strong magnetic fields would decay on a timescale of $\tau_B\sim10^4$ years. If their interpretation is correct, our result would be consistent with the one estimated from FRB observations. We therefore propose that the NSC-FRB systems can at least provide one significant channel for producing periodically repeating FRBs. As shown in Fig.~\ref{events}, the birth rate distribution for both NS+MS and NS+NS drop steeply for periods less than 10 days, inferring that if most periodically repeating FRBs were generated by the NSC-FRB systems, there should be very less FRBs with periods less than 10 days. Very interestingly, considering that the funnel would be spiraled by the orbital motion within the photosphere, \cite{ioka20} also predicted a lack of FRBs with periods less than 10 days for NS+MS systems and less than 0.1 days for NS+NS systems. These two results from different perspectives confirm the lack of periodic FRBs with short periods.

\section{Conclusion and discussion}
\label{s_conclusion}

The recent discovery of the periodically repeating FRB 180916 may bring clues for the source of repeating FRBs. The binary system containing an NS and its companion star (e.g. NSC-FRB system) has been well discussed to be the source of periodically repeating FRBs. Following the ``Binary Comb" model proposed by \cite{ioka20}, in this \emph{Letter} we use the population synthesis method to study in detail the properties of the companion stars and the nature of NSC-FRB system. Our analysis shows that for NSC-FRB systems, the companion star is likely to be a B-type star with orbital eccentricity in a wide distribution from 0.20-0.90. Note that here we did not consider the quick circularization of the orbit. On the one hand, the circularization effect might reduce the upper boundary of the orbital eccentricity, on the other hand, it may reduce the number of sources that meet the optical depth criterion ($\tau_{\rm C}<10$), because the neutron star becomes closer to the companion star.

The period of 16 days of FRB 180916 happens to fall in the most probable period range, which may explain why FRB 180916 was the first periodically repeating FRB to be detected, and we expect to observe more periodically repeating FRBs with periods around 10-30 days. When the companion is a B-type star, the birth rate for NSC-FRB systems with 15-17 days is estimated as $5.5 \times 10^{2}\,\rm Gpc^{-3} yr^{-1}$. This value will be affected by the initial parameters for binary star evolution. \citet{Zapartas2017} applied similar initial parameters for calculating neutron star formation, and they indicated that the change of total number events of CCSN with a variety of settings is less than $25\%$. Thus, as a very brief estimation, we would expect the change in the birth rate in this work is around $25\%$, either. In this case, even if the observation effect, the restrictions for the magnetic field of the newly born NS and the orbital circularization effect are considered, the birth rate could still fulfill the event rate requirement set by the observation of FRB 180916, inferring that the NSC-FRB systems can at least provide one significant channel for producing periodically repeating FRBs. However, it is worth noting that in order to investigate in more depth the statistical properties of the NSC-FRB systems, a study of the sensitivity of the theoretical predictions (rates, distributions of properties) to the uncertain input physics needs to be investigated in the future work.

In summary, we propose that if the periodically repeating FRBs are indeed from NSC-FRB systems, in addition to discussing the formation mechanism of radio emission, it is also worth exploring the properties of the binary systems, which could be useful for future research on the observation and theoretical modeling for fast radio bursts. The specific results discussed in this \emph{Letter} are under the ``Binary Comb" model framework, however, the method proposed here can also be applied to other models as long as the criteria to become a NSC-FRB system being revised accordingly. For instance, \cite{lyutikov20} proposed that the observed periodicity is due to the orbital phase-dependent modulation of the absorption conditions in the massive star's wind, in this case, the NS needs to be strongly magnetized and the companion star wind density also needs to adjust to produce periodic FRBs.

\section*{Acknowledgments}
We thank the anonymous referee for the helpful comments that have helped us to improve the presentation of the paper. This work is supported by the grants No. 11722324, 11690024, 11703001, 10933002 and 11273007 from the National Natural Science Foundation of China, the Joint Research Fund in Astronomy (U1631236) under cooperative agreement between the National Natural Science Foundation of China (NSFC) and Chinese Academy of Sciences (CAS), the Strategic Priority Research Program of the Chinese Academy of Sciences, Grant No. XDB23040100  and the Fundamental Research Funds for the Central Universities.

\bibliographystyle{mn} % (uses file "abbrvna.bst")
\bibliography{mybib} % expects file "mybib.bib"

\label{lastpage}

\end{document}